\documentclass[a4paper, 11pt]{article}
\usepackage{latexsym}
\title{On Quantum Mechanics \\
on Noncommutative Quantum Phase Space}
\author{
A.E.F. DJEMA\"{I}\footnote{Permanent address : D\'epartement de
Physique, Facult\'e des Sciences, Universit\'e d'Oran Es--s\'enia,
31100, Oran, Algeria. Fax : (00).(213).(41).41.91.84,
e-mail : abedelfarid@yahoo.com} \\
\small{Abdus Salam International Centre for
Theoretical Physics, 34100, Trieste, Italy}\\
and \\
H. SMAIL \\
\small{D\'epartement de Physique, Institut d'Hydraulique,}\\
\small{Centre Universitaire Mustapha Stambouli, Mascara, 29000,
Algeria}}
\date{August, 2003}
\begin{document}
\maketitle
\begin{abstract}
\footnotesize{In this work, we develop a general framework in
which Noncommutative Quantum Mechanics (NCQM), characterized by a
space noncommutativity matrix parameter $\theta =
\epsilon_{ij}^{~~k}\theta_{k}$ and a momentum noncommutativity
matrix parameter $\beta_{ij} =\epsilon_{ij}^{~~k}\beta_{k}$, is
showed to be equivalent to Quantum Mechanics (QM) on a suitable
transformed Quantum Phase Space (QPS). Imposing some constraints
on this particular transformation, we firstly find that the
product of the two parameters $\theta$ and $\beta$ possesses a
lower bound in direct relation with Heisenberg incertitude
relations, and secondly that the two parameters are equivalent but
with opposite sign, up to a dimension factor depending on the
physical system under study. This means that
\textbf{noncommutativity} is represented by a unique parameter
which may play the role of a \textbf{fundamental constant}
characterizing the whole NCQPS. Within our framework, we treat
some physical systems on NCQPS : free particle, harmonic
oscillator, system of two--charged particles, Hydrogen atom. Among
the obtained results, we discover a new phenomenon which consists
to see a free particle on NCQPS as equivalent to a harmonic
oscillator with Larmor frequency depending on $\beta$,
representing the same particle in presence of a magnetic field
$\vec{B} = q^{-1}\vec{\beta}$. For the other examples, additional
correction terms depending on $\beta$ appear in the expression of
the energy spectrum. Finally, in the two--particle system case, we
emphasize the fact that for two opposite charges noncommutativity
is effectively perceived with opposite sign.}
\end{abstract}
\vspace{1truecm}
\begin{center}
\footnotesize{PACS : 03.65.ca and 11.10.Nx }
\end{center}
\newpage
\section{Introduction}
Let  $\bf{x_{i}}$, $\bf{p_{i}}$ be the position and momentum
operators which generate the Heisenberg algebra of QM:
\begin{eqnarray}
\left[ \bf{x_{i}} , \bf{x_{j}} \right]= 0,~~~~ \left[ \bf{x_{i}} ,
\bf{p_{j}} \right]= i \hbar \delta_{ij}\textbf{1},~~~~
\left[\bf{p_{i}}, \bf{p_{j}}\right]= 0      \label{a}
\end{eqnarray}
where the dynamics is described by the canonical equations :
\begin{eqnarray}
\bf{\dot{x}}_{i}= \left[ \bf{x_{i}} , \bf{H} \right]~~~~,~~~~
\bf{\dot{p}_{i}}= \left[ \bf{p_{i}}, \bf{H} \right]  \label{b}
\end{eqnarray}
with $\bf{H}$ being the Hamiltonian operator of the quantum system
described on QPS.\\
In the context of "Quantization by deformation", \cite{1}, it has
been shown that this operator algebra is equivalent to a $\hbar
$-star deformation of the Poisson algebra of classical observables
equipped with a Weyl--Wigner--Moyal product defined as follows:
\begin{eqnarray*}
(f\ast_{\hbar } g)(u) = exp\left[ \frac {i\hbar }
{2}\omega^{ab}\partial^{(1)}_{a}\partial^{(2)}_{b} \right]
f(u_{1})g(u_{2})|_{u_{1}= u_{2}= u}
\end{eqnarray*}
where $u_{a}$ are phase space variables, $a=1,....,2N$, and $\omega$
is the usual symplectic structure .\\
Moreover, it has been shown in \cite{2} that QM can be formulated
as a noncommutative symplectic geometry by means of a discrete
Weyl--Schwinger realization of the Heisenberg group and by
developing a discrete version of the Weyl--Wigner-Moyal formalism.
In analogy with the classical case, the noncommutative (quantum)
symplectic geometry associated with the matrix algebra
$M_{N}(\textbf{C})$ generated by Schwinger basis was described.\\
Recently, there has been a growing interest on the description of
QM on non-commutative spaces (NCQM), \cite{3}---\cite{6}. This was
motivated by string theory arguments, which tell us that the low
energy effective theory of D--brane in the background of NS--NS
$B$ field lives on noncommutative spaces, \cite{7}. In fact, in
this context, our space--time may be the worldvolume of a
D--brane, and thus may be noncommutative. Moreover, it has been
shown that the study of the dynamics of a quantum system on a
noncommutative space is equivalent to the study of the dynamics of
this system on a commutative space in
presence of a magnetic field $\vec{B}$, \cite{5}.\\
A \textbf{noncommutative space} can be realized by coordinate
operators satisfying :
\begin{eqnarray*}
[ \textbf{x}_{\mu} , \textbf{x}_{\nu} ]_{\star} = i \hbar
\theta_{\mu \nu}
\end{eqnarray*}
where $\theta_{\mu\nu}$ is the noncommutativity parameter of
dimension $\frac{(lenght)^{2}}{\hbar}$ represented by an
antisymmetric matrix whose entries $\theta_{0i}$ are considered to
be vanishing, otherwise the theory is not unitary. Performing
explicit loop calculations, it has been proved that, for instance,
the NonCommutative QED (NCQED) up to one loop is renormalizable.
Then, in order to study phenomenological consequences of the
noncommutativity of the space, it is more indicated to try to
learn more about the QM on NC spaces.\\
The aim of this work is to study QM on NCQPS. We find that, in
addition to the results obtained in various papers where the
noncommutativity is only present in the space sector, there are
additional terms due to the noncommutativity $\beta$ present in
the momentum sector of QPS since its existence is, in fact, due
essentially to the existence of the noncommutativity $\theta$ on the space. \\
In this work, we construct a general $\alpha$--star deformation of
the algebra of classical observables that gives rise to a general
NCQM. It appears that this $\alpha$--star deformation is
equivalent to some general transformation on the usual quantum
phase space variables $(\textbf{x}_{i}, \textbf{p}_{i})$. Indeed,
we show that NCQM
is equivalent to QM on a transformed QPS.\\
This paper is organized as follows. In section 2, we describe a
general NCQM. The section 3 is devoted to study the noncommutative
Hamiltonian dynamics of a quantum system. Some simple examples of
quantum systems are considered like : free particle, harmonic
oscillator, two--particle system and as a particular case the
Hydrogen atom. Finally, section 4 is devoted to some conclusions
and perspectives.
\section{Description of a general Noncommutative quantum \\
Mechanics}
Let us consider a general $\alpha $-star deformation on the
Poisson algebra of the classical observables  as:
\begin{eqnarray*}
(f\ast_{\alpha} g)(u)= exp\left[
\frac{1}{2}\alpha_{ab}\partial_{a}^{(1)}\partial_{b}^{(2)}\right]
f(u_{1})g(u_{2})|_{u_{1}=u_{2}=u}
\end{eqnarray*}
where $\alpha $ generalizes the usual classical symplectic
structure $\omega$ and its general form may be given by:
\begin{eqnarray}
\alpha = \left(
          \begin{array}{cc}
          \theta               & \textbf{1} + \sigma\\
          -\textbf{1} - \sigma & \beta
          \end{array}
          \right)  \label{x}
\end{eqnarray}
where $\theta$ and $\beta$ are antisymmetric $3\times 3$  matrices
and $\sigma$ a symmetric $3\times 3$  matrix, such that:
\begin{eqnarray*}
\theta_{ij} = \epsilon _{ij}^{~~k}\theta_{k}~~~~,~~~~
\beta_{ij}=\epsilon_{ij}^{~~k}\beta_{k}
\end{eqnarray*}
The $\alpha$-star deformed Poisson algebra is given by:
\begin{eqnarray*}
\{ x_{i}, x_{j}\}_{\alpha } = \theta_{ij}~~,~~ \{ x_{i}, p_{j}
\}_{\alpha} = \delta_{ij}+ \sigma_{ij}~~,~~ \{ p_{i} , p_{j}
\}_{\alpha} = \beta_{ij}.
\end{eqnarray*}
It is easy to see that the \textit{classical limit} is guaranteed
by the condition $\theta = \beta = \sigma = \textbf{0}$.\\
Considering the noncommutativity as a small perturbation on the
structure of the phase space, the real parameters $\theta_{i}$ and
$\beta_{i}$ are taken very small and our calculations are taken up
to first order in $\theta$ and $\beta$. We will see that $\sigma$
can be ignored since it is of second order.\\
The above star-product permits us to deduce the star deformed
Heisenberg algebra which defines a general NCQM:
\begin{eqnarray}
\left[ \bf{x_{i}}, \bf{x_{j}}\right]_{\alpha } =
i\hbar\theta_{ij}\textbf{1}~~,~~ \left[ \bf{x_{i}},\bf{p_{j}}
\right]_{\alpha} = i\hbar (\delta_{ij}+
\sigma_{ij})\textbf{1}~~,~~ \left[ \bf{p_{i}} , \bf{p_{j}}
\right]_{\alpha} = i\hbar\beta_{ij} \textbf{1} \label{1}
\end{eqnarray}
via a generalized Dirac quantization :
\begin{eqnarray*}
\{ , \}_{\alpha} \longrightarrow \frac{1}{i\hbar} [ , ]_{\alpha}
\end{eqnarray*}
and with a noncommutative quantum dynamics governed by the motion
equations :
\begin{equation}
\dot{\textbf{x}}_{i} = \left[ \textbf{x}_{i} , \textbf{H}
\right]_{\alpha}~~~~,~~~~\dot{\textbf{p}_{i}} = \left[
\textbf{p}_{i} , \textbf{H} \right]_{\alpha} \label{z}
\end{equation}
Here, we argue that introducing a noncommutativity parametrized by
$\theta$ on the space sector of the QPS, we automatically
introduce a noncommutativity parametrized by another parameter,
say $\beta$, on the momentum sector of QPS, since they are linked
by the famous Heisenberg incertitude relations, and then, the
parameter $\sigma$ which appears in the star--commutator of a
space operator with a momentum operator must be tied to the two
parameters $\theta$ and $\beta$. Hereunder, we will give a more
precise explanation to this point.\\
Let us consider a general linear transformation on the usual
quantum phase space variables:
\begin{eqnarray*}
  \left\{
        \begin{array}{l}
        {\textbf{x}'_{i}} = a_{ik} \textbf{x}_{k}+ b_{ik}
        \textbf{p}_{k}\\
        \textbf{p}'_{i} = c_{ik} \textbf{x}_{k} + d_{ik} \textbf{p}_{k}
        \end{array}
\right.
\end{eqnarray*}
At this stage, we remark that for the following particular choice
of the matrix parameters $a, b, c,$ and $d$:
\begin{eqnarray*}
 a = d =\textbf{1}~~~,~~~c = \frac{1}{2}\beta~~~,~~~b = -\frac{1}{2}\theta
\end{eqnarray*}
the new quantum variables $\textbf{x}'_{i}$ and $\textbf{p}'_{i}$:
\begin{eqnarray}
\left( \begin{array}{c}
          \textbf{x}'_{i}\\
          \textbf{p}'_{i}
       \end{array}
\right) = \left( \begin{array}{cc}
\delta_{ik}           & - \frac{1}{2}\theta_{ik} \\
\frac{1}{2}\beta_{ik} & \delta_{ik}
\end{array} \right)
\left( \begin{array}{c}
          \textbf{x}_{k}\\
          \textbf{p}_{k}
       \end{array}
\right) = T_{ik} \left(\begin{array}{c}
                         \textbf{x}_{k}\\
                         \textbf{p}_{k}
                       \end{array}
\right)
 \label{2}
\end{eqnarray}
or, in another manner:
\begin{eqnarray*}
\left\{
        \begin{array}{l}
        \vec \textbf{x}' = \vec \textbf{x}-\frac{1}{2} \vec \textbf{p}\wedge \vec \theta =
        \vec \textbf{x}-\frac{1}{2} \theta \vec \textbf{p} \\
        \vec \textbf{p}' = \vec \textbf{p} + \frac{1}{2}\vec
        \textbf{x}\wedge \vec \beta = \vec \textbf{p} + \frac{1}{2} \beta
        \vec \textbf{x}
        \end{array}
 \right.
\end{eqnarray*}
satisfy commutation relations that look like (\ref{1}) :
\begin{eqnarray}
\left[ \textbf{x}'_{i}, \textbf{x}'_{j}\right] =
i\hbar\theta_{ij}\textbf{1}~~,~~ \left[ \textbf{x}'_{i},
\textbf{p}'_{j}\right] = i\hbar(\delta_{ij}+
\sigma_{ij})\textbf{1}~~,~~ \left[ \textbf{p}'_{i},
\textbf{p}'_{j}\right]= i\hbar\beta_{ij}\textbf{1} \label{3}
\end{eqnarray}
where $\sigma$ must satisfy the following condition :
\begin{eqnarray*}
\sigma = -\frac{1}{8}(\theta\beta + \beta\theta)
\end{eqnarray*}
and so, it can be ignored since we are only interested by first
order terms in $\theta$ and $\beta$.\\
Then, instead of using quantum variables with
$\ast_{\alpha}$-product (see (\ref{1})), it is more suitable to
use the transformed quantum variables $(\textbf{x}'_{i},
\textbf{p}'_{i})$ with the usual operator product.\\
Investigating the transformation (\ref{2} ), we see that its
Jacobian is given by:
\begin{eqnarray*}
J = det(T) = (1+\frac{1}{8}\rho)^{2} \neq 0
\end{eqnarray*}
where:
\begin{eqnarray*}
\rho = tr(\theta .\beta) = tr(\beta .\theta) = -2\vec
\theta.\vec\beta
\end{eqnarray*}
Imposing to $J$ to be equal to 1, the parameter $\rho$ is
constrained to fulfill the following condition:
\begin{eqnarray}
\rho = -16 \Longrightarrow \vec \theta .\vec\beta = 8 \label{y}
\end{eqnarray}
From the relations (\ref{3}), we see that the dimension of
$\theta_{i}$ is the same as $\frac{(\triangle x_{i})^{2}}{\hbar}$
and the dimension of $\beta_{i}$ as $\frac{(\triangle
p_{i})^{2}}{\hbar}$ and then the above condition looks like the
well known Heisenberg incertitude relations :
\begin{eqnarray*}
\Delta x_{i}.\Delta p_{i}  \geq  \hbar
\end{eqnarray*}
It follows that the noncommutative perturbation along the
space-like and momentum-like sectors must verify the same kind of
incertitude relations as do the position and the momentum
operators. The right side of the equation (\ref{y}) gives in fact
the lower bound of the product of the two parameters $\theta$ and $\beta$.\\
Furthermore, imposing to the matrix transformation $T$ to be
orthogonal :
\begin{eqnarray*}
TT^{t} = T^{t}T = \textbf{1}
\end{eqnarray*}
we deduce that the deformation parameters $\theta_{i}$ in the
space-like sector of the quantum phase space are of the same
"magnitude", \textbf{up to a dimension factor depending on the
physical system under study}, as the deformation parameters
$\beta_{i}$ in the momentum-like sector:
\begin{eqnarray*}
\theta_{i}\sim -\beta_{i}
\end{eqnarray*}
Effectively, this special orthogonal transformation on the quantum
phase space expresses that, if one introduces a noncommutativity
as a perturbation on the space like sector of QPS, this leads
automatically to introduce a noncommutativity as an opposite
equivalent perturbation on the momentum-like sector. In fact, $T$
looks like an infinitesimal transformation on QPS :
\begin{eqnarray*}
T = \left( \begin{array}{cc}
            \textbf{1} & \textbf{0} \\
            \textbf{0} & \textbf{1}
            \end{array}
     \right) - \frac{1}{2} \left( \begin{array}{cc}
                                   \textbf{0} & \theta \\
                                   - \beta    & \textbf{0}
                                   \end{array}
                            \right)
\end{eqnarray*}
Hereunder, we will show that noncommutative effects are always
tied to the angular momentum of the system under study and so,
they emphasize the presence of some kind of \textbf{rotations}
relatively to the axis defined by $\vec{\theta}$.
\section{Noncommutative Quantum Dynamics}
In this section we investigate some examples of dynamical systems
on NCQPS within the framework presented in the previous section.
In particular, we propose to treat the cases of a free particle, a
harmonic oscillator, two--particle system and in particular
Hydrogen atom.\\
Let us begin to investigate the general case of a quantum system
with a Hamiltonian operator :
\begin{eqnarray*}
H(\vec{\textbf{x}}, \vec{\textbf{p}}) =
\frac{\vec{\textbf{p}}^{2}}{2m} + V(\vec{\textbf{x}})
\end{eqnarray*}
where $\vec\textbf{x}$ and $\vec\textbf{p}$ satisfy (\ref{a}). Its
quantum
dynamics is governed by the motion equation (\ref{b}).\\
We have shown that the study of this quantum system on a NCQPS
where the position and momentum operators obey the nontrivial
commutation relations (\ref{1}) is equivalent to study it on the
usual QPS subject to the transformation (\ref{2}), where the new
position and momentum operators $\textbf{x}'$ and $\textbf{p}'$
obey the commutation relations (\ref{3}).\\
In this context, the Hamiltonian operator varies as :
\begin{eqnarray}
\Delta H = H (\vec{\textbf{x}}' , \vec{\textbf{p}}' ) -
H(\vec{\textbf{x}}, \vec{\textbf{p}}) = -
\frac{1}{2m}\beta_{ik}\textbf{x}_{i}\textbf{p}_{k}+
[V(\vec{\textbf{x}}')-V(\vec{\textbf{x}})] = -
\frac{1}{2m}\vec{\textbf{L}}.\vec{\beta} + \Delta V  \label{e}
\end{eqnarray}
where $\vec{\textbf{L}} = \vec{\textbf{x}} \wedge
\vec{\textbf{p}}$ is the angular momentum of the quantum system.
It is clear that there exists a correction term $\Delta E_{NC}$ to
the energy spectrum of our quantum dynamical system due to the
presence of noncommutativity in the QPS reflected by the two
additional terms in the right side of the equation (\ref{e}).
\subsection{Free Particle}
In the case of a quantum free particle $(V(\vec{\textbf{x}})=
\textbf{0})$, the Hamiltonian transforms as (see (\ref{e})) :
\begin{eqnarray*}
H'(\vec{\textbf{x}},\vec{\textbf{p}}) =
\frac{\vec{\textbf{p}}'^{2}}{2m} = \frac{\vec{\textbf{p}}^{2}}{2m}
- \frac{1}{2m}\beta_{ik}\textbf{x}_{i}\textbf{p}_{k} =
\frac{\vec{\textbf{p}}^{2}}{2m} -
\frac{1}{2m}\vec{\textbf{L}}.\vec{\beta}
\end{eqnarray*}
where:
\begin{eqnarray}
\vec{\textbf{p}}' = \vec{\textbf{p}} + \frac{1}{2}(\vec
{\textbf{x}} \wedge \vec{\beta}) \label{d}
\end{eqnarray}
This shows that the study of the dynamics of a particle on a NCQPS
is equivalent to the study of the dynamics of this particle of
charge $q$ on the usual QPS in presence of a magnetic field. This
interpretation  is justified by the identification of the
additional term in (\ref{d}) as a vector potential $\vec
{\textbf{A}}$ associated with a magnetic field $\vec {B}$:
\begin{eqnarray*}
q\vec {\textbf{A}} = -\frac{1}{2}
\vec{\textbf{x}}\wedge\vec{\beta} ~~~~~~/~~~~~~ \vec{\beta} = q
\vec{B}
\end{eqnarray*}
Considering the simple case $\beta_{1} = \beta_{2} = 0$ and
$\beta_{3} = \beta = qB$, and defining :
\begin{eqnarray*}
\textbf{z} = \sqrt{\frac{|\beta|}{2\hbar}} \left( \textbf{x} +
i\textbf{y} \right)
\end{eqnarray*}
and the annihilation and creation operators :
\begin{eqnarray*}
\textbf{a} = \partial_{\overline{z}} +
\frac{1}{2}\textbf{z}~~~,~~~\textbf{a}^{+} = -
\partial_{z} + \frac{1}{2} \overline{\textbf{z}}
\end{eqnarray*}
one finds that our quantum particle looks like a harmonic
oscillator system :
\begin{eqnarray*}
[ \textbf{a} , \textbf{a} ] = [ \textbf{a}^{+} , \textbf{a}^{+} ]
= \textbf{0}~~~,~~~[ \textbf{a} , \textbf{a}^{+} ] = \textbf{1}
\end{eqnarray*}
with Hamiltonian operator :
\begin{eqnarray*}
H( \textbf{a} , \textbf{a}^{+}) = \hbar \omega_{\beta} \left(
\textbf{a}^{+}\textbf{a} + \frac{1}{2}\textbf{1} \right)
\end{eqnarray*}
giving us a spectrum formed by infinite degenerate Landau levels :
\begin{eqnarray*}
E_{l} = \hbar \omega_{\beta} \left(  l + \frac{1}{2}\right)~~~~~,
~~l \in \mathbf{Z}
\end{eqnarray*}
where
\begin{eqnarray*}
\omega_{\beta} = \frac{|\beta|}{m} = \frac{|qB|}{m}
\end{eqnarray*}
denotes the Larmor frequency.\\
Finally, we conclude that introducing a noncommutativity on the
QPS, the free quantum particle behaves as a harmonic oscillator
with a (very small) frequency depending on the noncommutative
perturbation $\beta$ on the momentum sector.
\subsection{Harmonic oscillator}
Let us consider now the example of a quantum harmonic oscillator
of charge $q$ and with a potential :
\begin{eqnarray*}
V(\vec{\textbf{x}}) = \frac{1}{2} \kappa \vec{\textbf{x}}^{2}
\end{eqnarray*}
and let set :
\begin{eqnarray}
\vec{\mu} = \left[~ \vec{\beta} + m\kappa \vec{\theta} ~\right] .
\label{100}
\end{eqnarray}
Then, the modified Hamiltonian becomes (see (\ref{e})) :
\begin{eqnarray*}
H'(\vec{\textbf{x}}, \vec{\textbf{p}}) = H(\vec{\textbf{x}},
\vec{\textbf{p}}) - \frac{1}{2m}\vec{\textbf{L}}.\vec{\mu} =
\frac{\vec{\textbf{P}}^{2}}{2m} + V(\vec{\textbf{x}})
\end{eqnarray*}
where:
\begin{eqnarray*}
\vec{\textbf{P}} = \vec{\textbf{p}} + \frac{1}{2} \vec{\textbf{x}}
\wedge \vec{\mu}
\end{eqnarray*}
The resulting system is then always a quantum harmonic oscillator
but with a shifted energy spectrum. In fact, it looks also as the
ordinary harmonic oscillator in presence of a magnetic field :
\begin{eqnarray*}
\vec{B} = q^{-1} \vec{\mu} = q^{-1} \left[ ~\vec{\beta} +
m\kappa\vec{\theta}~\right]
\end{eqnarray*}
which is the sum of two magnetic fields $\vec{B}_{1} =
q^{-1}\vec{\beta}$ and $\vec{B}_{2} = q^{-1}m\kappa\vec{\theta}$
due to the presence of noncommutativity in the momentum and space
sectors of QPS
respectively.\\
The new Larmor frequency characterizing the correction term in the
system spectrum is :
\begin{eqnarray*}
\omega_{\mu} = \frac{\mid \mu \mid}{m} = \frac{ \mid q B
\mid}{m}~~.
\end{eqnarray*}
\subsection{Two--particle system}
Let us treat now a system of two distinct quantum particles with
respective masses and charges $(m_{a}, q_{a})$ and $(m_{b} ,
q_{b})$, defined on NCQPS. Moreover, since the dynamical states of
two different particles belong to two different Hilbert spaces,
one considers that their position and momentum operators commute
under any product, stared or not. This leads to the following set
of non--trivial commutation relations :
\begin{eqnarray}
[\hat{\textbf{x}}^{(a)}_{i} ,\hat{\textbf{x}}^{(a)}_{j} ~]_{\star}
= i\hbar \theta^{(a)}_{ij}~~,~~[\hat{\textbf{x}}^{(a)}_{i} ,
\hat{\textbf{p}}^{(a)}_{j} ~]_{\star} = i\hbar ( \delta_{ij} +
\sigma_{ij} )~~,~~[ \hat{\textbf{p}}^{(a)}_{i} ,
\hat{\textbf{p}}^{(a)}_{j} ~]_{\star} = i\hbar \beta^{(a)}_{ij}
\label{o}
\end{eqnarray}
\begin{eqnarray}
[\hat{\textbf{x}}^{(b)}_{i} ,\hat{\textbf{x}}^{(b)}_{j}~ ]_{\star}
= i\hbar \theta^{(b)}_{ij}~~,~~[\hat{\textbf{x}}^{(b)}_{i} ,
\hat{\textbf{p}}^{(b)}_{j}~]_{\star} = i\hbar ( \delta_{ij} +
\sigma_{ij} )~~,~~[\hat{\textbf{p}}^{(b)}_{i} ,
\hat{\textbf{p}}^{(b)}_{j}~ ]_{\star} = i\hbar \beta^{(b)}_{ij}
\label{k}
\end{eqnarray}
\begin{eqnarray}
[~ \hat{\textbf{x}}^{(a)}_{i} , \hat{\textbf{x}}^{(b)}_{j}
~]_{\star} = [~ \hat{\textbf{x}}^{(a)}_{i} ,
\hat{\textbf{p}}^{(b)}_{j} ~]_{\star} = [~
\hat{\textbf{p}}^{(a)}_{i} , \hat{\textbf{x}}^{(b)}_{j} ~]_{\star}
= [~ \hat{\textbf{p}}^{(a)}_{i} , \hat{\textbf{p}}^{(b)}_{j}
~]_{\star} = \textbf{0} \label{l}
\end{eqnarray}
Within our framework, the study of this two--particle system on
NCQPS is equivalent to study it on the usual QPS on which we
perform the transformation (\ref{2}), where $\sigma$ can be
ignored since it is of second order. This means that we have the
new variables :
\begin{eqnarray*}
\textbf{x'}^{(a)}_{i} = \textbf{x}^{(a)}_{i} -
\frac{1}{2}\theta^{(a)}_{ij}
\textbf{p}^{(a)}_{j}~~~,~~~\textbf{p'}^{(a)}_{i} =
\textbf{p}^{(a)}_{i} + \frac{1}{2}\beta^{(a)}_{ij}
\textbf{x}^{(a)}_{j}
\end{eqnarray*}
\begin{eqnarray*}
\textbf{x'}^{(b)}_{i} = \textbf{x}^{(b)}_{i} -
\frac{1}{2}\theta^{(b)}_{ij}
\textbf{p}^{(b)}_{j}~~~,~~~\textbf{p'}^{(b)}_{i} =
\textbf{p}^{(b)}_{i} + \frac{1}{2}\beta^{(b)}_{ij}
\textbf{x}^{(b)}_{j}
\end{eqnarray*}
where the "primed" variables obey the same commutation relations
(\ref{o}),(\ref{k}) and (\ref{l}) as do the noncommutative
variables, but without $\star$--product. The non--primed variables
$\textbf{x}^{(a,b)}_{i}$ and $\textbf{p}^{(a,b)}_{i}$ generate
usual
Heisenberg algebras for each particle.\\
To deal with the two--particle system, let us consider the
following more convenient set of operators :
\begin{eqnarray*}
\textbf{X}_{i} &=& \textbf{x}^{(a)}_{i} -\textbf{x}^{(b)}_{i}
~~~~~~~~~~~~~~"relative~coordinate~operators" \\
\textbf{Y}_{i} &=& \frac{m_{a}\textbf{x}^{(a)}_{i} +
m_{b}\textbf{x}^{(b)}_{i}}{m_{a} + m_{b}} ~~~~~~"center~ of~ mass~
coordinate~operators" \\
\textbf{P}_{i} &=& \textbf{p}^{(a)}_{i} +
\textbf{p}^{(b)}~~~~~~~~~~~~~~"total~momentum~operators" \\
\textbf{Q}_{i} &=& \frac{m_{b}\textbf{p}^{(a)}_{i} -
m_{a}\textbf{p}^{(b)}_{i}}{m_{a} + m_{b}}~~~~~~"relative~momentum
~operators"
\end{eqnarray*}
and let us introduce :
\begin{eqnarray*}
M &=& m_{a} + m_{b}~~~~~~~~~~~~~~"total~mass" \\
\mu &=& \frac{m_{a} m_{b}}{m_{a} + m_{b}}~~~~~~~~~~"reduced~ mass"
\end{eqnarray*}
These new two--particle system operators verify the following set
of commutation relations:
\begin{eqnarray*}
\left[  \textbf{X}_{i} , \textbf{Q}_{j}  \right] = \left[
\textbf{Y}_{i} , \textbf{P}_{j} \right] = i\hbar \delta_{ij}
\end{eqnarray*}
and all the others are vanishing.\\
The primed corresponding operators satisfy :
\begin{eqnarray}
\left[ {\textbf{X}'}_{i} , {\textbf{X}'}_{j} \right] &=& i\hbar
\left( \theta^{(a)}_{ij} + \theta^{(b)}_{ij}
\right)\textbf{1}~~,~~\left[ {\textbf{X}'}_{i} , {\textbf{Y}'}_{j}
\right] = i\hbar \left( \frac{m_{a}}{M}\theta^{(a)}_{ij} -
\frac{m_{b}}{M}\theta^{(b)}_{ij} \right)\textbf{1}
\nonumber \\
\left[ {\textbf{X}'}_{i} , {\textbf{P}'}_{j} \right]
&=& \textbf{0}~~,~~\left[ {\textbf{X}'}_{i} ,
{\textbf{Q}'}_{j} \right] = i\hbar \delta_{ij}\textbf{1} \nonumber \\
 \left[
{\textbf{Y}'}_{i} , {\textbf{Y}'}_{j} \right] &=& i\hbar \left(
\frac{m^{2}_{a}}{M^{2}} \theta^{(a)}_{ij} +
\frac{m^{2}_{b}}{M^{2}} \theta^{(b)}_{ij}
\right)\textbf{1}~~,~~\left[ {\textbf{Y}'}_{i} , {\textbf{P}'}_{j}
\right] = i\hbar \delta_{ij}\textbf{1} \nonumber \\
\left[
{\textbf{Y}'}_{i} , {\textbf{Q}'}_{j} \right] &=&
\textbf{0}~~,~~\left[ {\textbf{P}'}_{i} , {\textbf{P}'}_{j}
\right] = i\hbar \left( \beta^{(a)}_{ij} +
\beta^{(b)}_{ij} \right) \textbf{1} \nonumber \\
\left[ {\textbf{P}'}_{i} , {\textbf{Q}'}_{j} \right] &=& i\hbar
\left( \frac{m_{b}}{M} \beta^{(a)}_{ij} - \frac{m_{a}}{M}
\beta^{(b)}_{ij}\right) \textbf{1}~~,~~\left[ {\textbf{Q}'}_{i} ,
{\textbf{Q}'}_{j} \right] = i\hbar \left(
\frac{m^{2}_{b}}{M^{2}}\beta^{(a)}_{ij} + \frac{m^{2}_{a}}{M^{2}}
\beta^{(b)}_{ij} \right) \textbf{1} \label{11}
\end{eqnarray}
Under our transformation, the Hamiltonian of this two--particle
system :
\begin{eqnarray*}
H \left( \vec{\textbf{x}}^{(a)} , \vec{\textbf{x}}^{(b)} ,
\vec{\textbf{p}}^{(a)} , \vec{\textbf{p}}^{(b)} \right) =
\frac{\textbf{p}^{(a)}_{i} \textbf{p}^{(a)i}}{2m_{a}} +
\frac{\textbf{p}^{(b)}_{i} \textbf{p}^{(b)i}}{2m_{b}} + V \left(
\vec{\textbf{x}}^{(a)},\vec{\textbf{x}}^{(b)} \right)
\end{eqnarray*}
or, in terms of the new (non--primed) variables :
\begin{eqnarray*}
H \left( \vec{\textbf{X}} , \vec{\textbf{Y}} , \vec{\textbf{P}} ,
\vec{\textbf{Q}} \right) = \frac{\textbf{P}_{i}\textbf{P}^{i}}{2M}
+ \frac{\textbf{Q}_{i}\textbf{Q}^{i}}{2\mu} + V\left(
\vec{\textbf{X}} , \vec{\textbf{Y}} \right)
\end{eqnarray*}
will change like :
\begin{eqnarray*}
H \left( \vec{\textbf{x}'}^{(a)} , \vec{\textbf{x}'}^{(b)} ,
\vec{\textbf{p}'}^{(a)} , \vec{\textbf{p}'}^{(b)} \right) =
\frac{{\textbf{p}'}^{(a)}_{i} {\textbf{p}'}^{(a)i}}{2m_{a}} +
\frac{{\textbf{p}'}^{(b)}_{i} {\textbf{p}'}^{(b)i}}{2m_{b}} +
V\left( \vec{\textbf{x}'}^{(a)},\vec{\textbf{x}'}^{(b)} \right)
\end{eqnarray*}
or equivalently :
\begin{eqnarray}
H \left( \vec{\textbf{X}}' , \vec{\textbf{Y}}' , \vec{\textbf{P}}'
, \vec{\textbf{Q}}' \right) =
\frac{{\textbf{P}'}_{i}{\textbf{P}'}^{i}}{2M} +
\frac{{\textbf{Q}'}_{i}{\textbf{Q}'}^{i}}{2\mu} + V\left(
\vec{\textbf{X}}' , \vec{\textbf{Y}}' \right) \label{12}
\end{eqnarray}
where :
\begin{eqnarray}
{\textbf{X}'}_{i} &=& \textbf{X}_{i} - \frac{1}{2}
\theta^{(a)}_{ij} \textbf{p}^{(a)}_{j} + \frac{1}{2}
\theta^{(b)}_{ij} \textbf{p}^{(b)}_{j}\nonumber \\
{\textbf{Y}'}_{i} &=& \textbf{Y}_{i} - \frac{m_{a}}{2M}
\theta^{(a)}_{ij} \textbf{p}^{(a)}_{j} -
\frac{m_{b}}{2M} \theta^{(b)}_{ij} \textbf{p}^{(b)}_{j} \nonumber \\
{\textbf{P}'}_{i} &=& \textbf{P}_{i} + \frac{1}{2}\beta^{(a)}_{ij}
\textbf{x}^{(a)}_{j} + \frac{1}{2}\beta^{(b)}_{ij}
\textbf{x}^{(b)}_{j} \nonumber \\
{\textbf{Q}'}_{i} &=&
\textbf{Q}_{i} + \frac{m_{b}}{2M}\beta^{(a)}_{ij}
\textbf{x}^{(a)}_{j} - \frac{m_{a}}{2M}\beta^{(b)}_{ij}
\textbf{x}^{(b)}_{j}  \label{10}
\end{eqnarray}
At this level, we recall our assumption presented in the precedent
section which consists to consider that, \textbf{up to a dimension
factor depending on the physical system under study}, we have for
the same particle :
\begin{eqnarray*}
\beta^{(a)}_{ij} = - \theta^{(a)}_{ij}~~~~,~~~~\beta^{(b)}_{ij} =
- \theta^{(b)}_{ij}
\end{eqnarray*}
Furthermore, considering the particular and very interesting case
of two--particle system whose charges $q_{a}$ and $q_{b}$ have
opposite signs , and knowing that \textit{noncommutativity} means
the existence of a magnetic field on the QPS in presence of which
charges of opposite signs have opposite motions, we can also argue
that each of our two particles perceive the \textbf{same
noncommutativity but with opposite sign}, \cite{8} :
\begin{eqnarray*}
\theta^{(a)}_{ij} = - \theta^{(b)}_{ij} = \theta_{ij}~~~,~~~
\beta^{(a)}_{ij} = - \beta^{(b)}_{ij} = \beta_{ij}
\end{eqnarray*}
and so, finally, we are dealing with only one noncommutativity
parameter which characterizes the whole NCQPS :
\begin{eqnarray*}
\theta_{ij} = \theta^{(a)}_{ij} = - \theta^{(b)}_{ij} =
\beta^{(b)}_{ij} = - \beta^{(a)}_{ij}
\end{eqnarray*}
In this case, the relations (\ref{10}) reduce to :
\begin{eqnarray*}
{\textbf{X}'}_{i} &=& \textbf{X}_{i} -
\frac{1}{2}\theta_{ij}\textbf{P}_{j} \\
{\textbf{Y}'}_{i} &=& \textbf{Y}_{i} - \frac{1}{2}\theta_{ij}
\left[ \textbf{Q}_{j} + \left( \frac{m_{a} - m_{b}}{M}  \right)
\textbf{P}_{j} \right] \\
{\textbf{P}'}_{i} &=& \textbf{P}_{i} - \frac{1}{2} \theta_{ij}
\textbf{X}_{j} \\
{\textbf{Q}'}_{i} &=& \textbf{Q}_{i} - \frac{1}{2} \theta_{ij}
\left[ \textbf{Y}_{j} - \left( \frac{m_{a} - m_{b}}{M} \right)
\textbf{X}_{j} \right]
\end{eqnarray*}
and the relations (\ref{11}) become :
\begin{eqnarray*}
\left[ {\textbf{X}'}_{i} , {\textbf{Y}'}_{j} \right] &=& - \left[
{\textbf{P}'}_{i} , {\textbf{Q}'}_{j} \right] = i\hbar \theta_{ij}
\textbf{1} \\
\left[ {\textbf{X}'}_{i} , {\textbf{Q}'}_{j} \right] &=& \left[
{\textbf{Y}'}_{i} , {\textbf{P}'}_{j} \right] = i\hbar \delta_{ij}
\textbf{1} \\
\left[ {\textbf{Y}'}_{i} , {\textbf{Y}'}_{j} \right] &=& \left[
{\textbf{Q}'}_{i} , {\textbf{Q}'}_{j} \right] = i\hbar \left(
\frac{m_{a} - m_{b}}{M} \right)\theta_{ij}\textbf{1} \\
\left[ {\textbf{X}'}_{i} , {\textbf{X}'}_{j} \right] &=& \left[
{\textbf{X}'}_{i} , {\textbf{P}'}_{j} \right] = \left[
{\textbf{Y}'}_{i} , {\textbf{Q}'}_{j} \right] = \left[
{\textbf{P}'}_{i} , {\textbf{P}'}_{j} \right] = \textbf{0}
\end{eqnarray*}
With these simplifications, the transformed Hamiltonian (\ref{12})
will differ from the non--transformed one by :
\begin{eqnarray*}
\Delta H &=& - \frac{1}{2M} \theta_{ij} \textbf{X}_{j}
\textbf{P}_{i} - \frac{1}{2\mu}
\theta_{ij}\textbf{Q}_{i}\textbf{Y}_{j} - \left( \frac{m_{b} -
m_{a}}{2m_{a}m_{b}}\right) \theta_{ij}\textbf{Q}_{i}\textbf{X}_{j}
+ V ( \textbf{X}' , \textbf{Y}' ) - V ( \textbf{X} , \textbf{Y} )
\\
       &=& \left[ \frac{1}{2M}(\vec{\textbf{X}} \wedge \vec{\textbf{P}})
+ \frac{1}{2\mu} (\vec{\textbf{Y}} \wedge \vec{\textbf{Q}}) +
\left( \frac{m_{b} - m_{a}}{2m_{a}m_{b}} \right) (\vec{\textbf{X}}
\wedge \vec{\textbf{Q}} ) \right]. \vec{\theta} + \Delta V
\end{eqnarray*}
which reduces to :
\begin{eqnarray}
\Delta H = \left[ \frac{1}{2m_{a}}\vec{\textbf{L}}^{(a)} -
\frac{1}{2m_{b}} \vec{\textbf{L}}^{(b)}\right].\vec{\theta} +
\Delta V    \label{p}
\end{eqnarray}
where :
\begin{eqnarray*}
\vec{\textbf{L}}^{(a)} = \vec{\textbf{x}}^{(a)} \wedge
\vec{\textbf{p}}^{(a)}~~~,~~~\vec{\textbf{L}}^{(b)} =
\vec{\textbf{x}}^{(b)} \wedge \vec{\textbf{p}}^{(b)}
\end{eqnarray*}
are the usual momentum operators of the two particles "a" and
"b".\\
Let us discuss now the dynamical states of the two--particle
system represented by a \textit{collective wave function} solution
of the \textbf{NC Schr\"{o}dinger equation}, \cite{3}, \cite{5} :
\begin{eqnarray*}
i\hbar \frac{\partial}{\partial t}\Psi(~\vec{x}^{(a)},
\vec{x}^{(b)} , t ) = \left[ - \frac{\hbar^{2}}{2m_{a}}
\Delta^{(a)} - \frac{\hbar^{2}}{2m_{b}} \Delta^{(b)} +
V(~\vec{x}^{(a)} , \vec{x}^{(b)} ) \right] \star \Psi(
\vec{x}^{(a)} , \vec{x}^{(b)} , t)
\end{eqnarray*}
where the generalized $\star$--product is defined by :
\begin{small}
\begin{eqnarray*}
(f \star g)(\vec{x}^{(a)} , \vec{x}^{(b)} ) = \exp \left\{
i\frac{\hbar}{2} \left( \theta^{(a)}_{ij} \frac{\partial}{\partial
y^{(a)}_{i}} \frac{\partial}{\partial z^{(a)}_{j}}  +
\theta^{(b)}_{ij} \frac{\partial}{\partial y^{(b)}_{i}}
\frac{\partial}{\partial z^{(b)}_{j}} \right) \right\} \times \\
f(\vec{y}^{(a)} , \vec{y}^{(b)}) g( \vec{z}^{(a)} ,
\vec{z}^{(b)})|_{ \left\{\begin{array}{c}
                      \vec{y}^{(a)}=\vec{z}^{(a)} =
                      \vec{x}^{(a)}\\
                      \vec{y}^{(b)}=\vec{z}^{(b)} = \vec{x}^{(b)}
                 \end{array} \right. }
\end{eqnarray*}
\end{small}
Let us recall that the above relations concern "hated" variables.
Instead to use these variables and following our approach, it is
more suitable to use our "primed" variables which verify the same
commutation relations as do the first ones but without
$\star$--product. It follows that the NC Schr\"{o}dinger equation
reads now as :
\begin{eqnarray*}
i\hbar \frac{\partial}{\partial t}\Psi(~\vec{x'}^{(a)},
\vec{x'}^{(b)} , t ) = \left[ - \frac{\hbar^{2}}{2m_{a}}
\Delta'^{(a)} - \frac{\hbar^{2}}{2m_{b}} \Delta'^{(b)} +
V(~\vec{x'}^{(a)} , \vec{x'}^{(b)} ) \right]  \Psi( \vec{x'}^{(a)}
, \vec{x'}^{(b)} , t)
\end{eqnarray*}
or equivalently,
\begin{eqnarray*}
i\hbar \frac{\partial}{\partial t}\Psi(~\vec{X'}, \vec{Y'} , t ) =
\left[ - \frac{\hbar^{2}}{2\mu} \Delta_{X'} - \frac{\hbar^{2}}{2M}
\Delta_{Y'} + V(~\vec{X'} , \vec{Y'}) \right]  \Psi( \vec{X'} ,
\vec{Y'} , t)
\end{eqnarray*}
To solve this equation, we use the technique of separation of
variables :
\begin{eqnarray*}
\Psi(X' , Y' , t) = \varphi(X')\phi(Y')\exp[-i\frac{E'}{\hbar}t]
\end{eqnarray*}
It results two separated differential equations concerning the
functions $\varphi(X')$ and $\phi(Y')$. Solving the second one,
one obtains :
\begin{eqnarray*}
\phi(Y') = \exp[i \vec{K'}.\vec{Y'} ]
\end{eqnarray*}
where $\vec{K}$ play the role of a wave vector associated to the
collective wave function relatively to the center of mass and
$\vec{K'}$ is its transformed. The second equation reads :
\begin{eqnarray*}
\Delta_{X'}\varphi(\vec{X'}) + \frac{2\mu}{\hbar^{2}}\left[ E' -
V(\vec{X'} , \vec{Y'}) - \frac{\hbar^{2}}{2M}\mid \vec{K'}\mid^{2}
\right] \varphi(\vec{X'}) = 0
\end{eqnarray*}
where :
\begin{eqnarray*}
{\vec{X}'} &=& \vec{X} + \frac{1}{2} \vec{\theta}\wedge \vec{P}\\
 {\vec{Y}'} &=& \vec{Y} + \frac{1}{2}\left[ \vec{\theta}\wedge
\vec{Q} + \left(\frac{m_{a} -
m_{b}}{M} \right) \vec{\theta} \wedge \vec{P} \right] \\
E' &=& E + \Delta E_{NC} \\
\Delta V &=& V(\vec{X'} , \vec{Y'}) -
V(\vec{X} , \vec{Y})
\end{eqnarray*}
Knowing that the usual two--particle Schr\"{o}dinger equation
reads :
\begin{eqnarray*}
\Delta_{X}\varphi(\vec{X}) + \frac{2\mu}{\hbar^{2}}\left[ E -
V(\vec{X} , \vec{Y}) - \frac{\hbar^{2}}{2M}\mid \vec{K}\mid^{2}
\right] \varphi(\vec{X}) = 0
\end{eqnarray*}
and using :
\begin{eqnarray*}
\Delta_{X'} \varphi(\vec{X'}) =  \left( \Delta_{X'} + \Delta_{P'}
\right)\varphi(\vec{X'}) = \left( \Delta_{X} + \Delta_{P}
\right)\varphi(\vec{X} + \frac{1}{2}\vec{\theta} \wedge \vec{P})
\simeq \Delta_{X} \varphi(\vec{X} + \frac{1}{2}\vec{\theta} \wedge
\vec{P})
\end{eqnarray*}
and considering the following separation of variables :
\begin{eqnarray*}
\varphi(\vec{X}') = \varphi(\vec{X}) F(\vec{\theta}\wedge \vec{P})
\end{eqnarray*}
we find that the noncommutative correction to the energy spectrum
of this two--particle system is given by :
\begin{eqnarray}
\Delta E_{NC} = \Delta V - \frac{\hbar^{2}}{2M} \left[
\mid\vec{K'} \mid^{2} - \mid\vec{K} \mid^{2} \right] .
\label{q}
\end{eqnarray}
In fact, it is easy to see that this relation is another
equivalent version of the relation (\ref{p}).
\subsection{Hydrogen atom}
This example can be treated in two different manners. The first
consists to consider this system as a one--particle system
(Electron) in an external Coulomb potential (Nucleus). In this
approach, we can directly use the relation (\ref{e}) to deduce the
NC correction of the Hamiltonian by injecting the potential :
\begin{eqnarray*}
V(\vec{\textbf{x}}) = -
\frac{Ze^{2}}{\sqrt{\textbf{x}_{i}\textbf{x}^{i}}} .
\end{eqnarray*}
The second method consists to consider it as a two--particle
system whose treatment is developed in the previous subsection. \\
Using the first method, we find that the variation of the
potential reads :
\begin{eqnarray*}
\Delta V = - \frac{Ze^{2}}{2\textbf{x}^{3}}
\vec{\textbf{L}}.\vec{\theta} = - \frac{e}{2}(\vec{\theta}\wedge
\vec{\textbf{p}}). \left( \frac{-Ze
\vec{\textbf{x}}}{\textbf{x}^{3}} \right)
\end{eqnarray*}
where $\textbf{x} = \sqrt{\textbf{x}_{i}\textbf{x}^{i}}$ and then,
the NC correction of the Hamiltonian is given by :
\begin{eqnarray*}
\Delta H = \frac{1}{2m} (\vec{\beta}\wedge \vec{\textbf{p}}).
\vec{\textbf{x}} - \frac{e}{2}(\vec{\theta}\wedge \vec{\textbf{p}}
). \left( \frac{-Ze\vec{\textbf{x}}}{\textbf{x}^{3}} \right) = -
\frac{1}{2m}\vec{\textbf{L}}.\vec{\eta}
\end{eqnarray*}
where
\begin{eqnarray}
\vec{\eta} = \vec{\beta} +
\frac{mZe^{2}}{\textbf{x}^{3}}\vec{\theta} . \label{200}
\end{eqnarray}
The first term in $\Delta H$ corresponds to the existence of a
magnetic field as already obtained in the precedent cases. It
comes as a kinetic correction term. The second one that comes as a
potential correction term looks like a spin--orbit coupling term
where the \textit{noncommutativity induced spin} momentum
$\vec{S}$ is given by  :
\begin{eqnarray*}
\vec{S} = \frac{\hbar}{\lambda_{e}^{2}}\vec{\theta}
\end{eqnarray*}
where $\lambda_{e}$ is the usual Compton wave length of the
electron, \cite{3}. \\
Following the second approach, we have first to
calculate $\Delta V$ (see (\ref{p})) :
\begin{eqnarray*}
\Delta V = - \frac{e}{2} (\vec{\theta} \wedge \vec{\textbf{P}}).
\left( \frac{-Ze^{2}\vec{\textbf{X}}}{\vec{\textbf{X}}} \right) =
- \frac{Ze^{2}}{2\textbf{X}^{3}}
\vec{\textbf{\textbf{\L}}}.\vec{\theta}
\end{eqnarray*}
where $\textbf{X} = \sqrt{\textbf{X}_{i}\textbf{X}^{i}}$ and
$\vec{\textbf{\textbf{\L}}} = \vec{\textbf{X}}\wedge \vec{\textbf{P}}$.\\
Then, the NC correction to the Hamiltonian is now given by :
\begin{eqnarray*}
\Delta H = \left[ \frac{1}{2m_{b}} \vec{\textbf{L}}^{(b)} -
\frac{1}{2m_{a}} \vec{\textbf{L}}^{(a)} \right]. \vec{\theta} -
\frac{Ze^{2}}{2\textbf{X}^{3}} \vec{\textbf{\L}}.\vec{\theta} .
\end{eqnarray*}
It is easy to see that this result reduces to the precedent one,
if one considers the nucleus (particle "b") as localized at the
origin and possessing an infinite mass.\\
Finally, in addition to the energy level shift at tree level for
Hydrogen atom obtained in \cite{3}, there exists another term
which takes into account the noncommutativity in the momentum
sector of NCQPS in such a way that :
\begin{eqnarray*}
\Delta E^{H--atom}_{NC} = - \frac{1}{2m}\langle
\vec{\textbf{L}}.\vec{\eta} \rangle = - \frac{1}{2m} \left[
\langle \vec{\textbf{L}}.\vec{\beta} \rangle + \langle
\frac{Ze^{2}}{2\textbf{X}^{3}}
\vec{\textbf{L}}.\vec{\theta}\rangle\right] .
\end{eqnarray*}
\section{Conclusion}
In \cite{4}, it is emphasized that there is \textbf{no
noncommutative corrections} at tree level for Hydrogen atom.
Solving the NC two--body Schr\"{o}dinger equation and considering
that $\theta_{ij}$ changes the sign under charge conjugation,
\cite{8}, the authors contradict the result obtained (and
confirmed) in \cite{3}.\\
In our work, following our approach, we have found that the
kinetic term in the Hamiltonian gives an additional correction
term to the result obtained in \cite{3}, (See (\ref{200})).
Following our assumption which consists to say that the
noncommutativity $\beta$ in the momentum sector of QPS is of the
same magnitude as the noncommutativity $\theta$ in the space
sector, up to a dimension factor depending on the physical system
under study, we are tempted to deduce that effectively there is no
noncommutative effects. We are motivated by the fact that
$\vec{\beta} \sim - \vec{\theta}$ and by the presence of a
coefficient of $\vec{\theta}$ in (\ref{200}) (as well as in
(\ref{100})), which is specific to the considered physical system.
It seems that this may ensure the annihilation of the two terms,
in such a way that the noncommutative effects do not really
appear. The proof that noncommutativity exists really is already
given by the simple example of free particle, which behaves like a
harmonic oscillator with a (very small) Larmor frequency tied to
$\beta$ (and so to $\theta$). It seems that it is the variation of
the potential when it exists which contributes
to the annihilation of this NC effect.\\
In another hand, it appears that \textbf{noncommutativity} is
deeply tied to the \textbf{presence of some magnetic sources} in
the space at scales near the planck one, and it is supposed to be
a \textbf{quantum effect of gravity}, \cite{9}. However, we think
that this problem needs a deep analysis to be well understood.\\
Furthermore, in \cite{10} an analog approach has been followed that
consists to contruct an \textit{isotropic representation} representing
general transformations on NCQPS. Nevertheless, in this approach the
authors have not used one important requirement on these transformations
which leads to to the natural condition $\theta = - \beta$ we have
obtained just by requiring an orthogonality property of these
transformations. In fact, their intention being to work in full generality,
they assumed that in general $\theta + \beta \neq 0$.\\
In future works, we plan to treat within our framework some other
quantum examples like : NCQED, Bohm--Aharanov, Lamb shift , Stark,
Zeeman and Hall effects,...\\
\textbf{Acknowledgments} : The authors would like to
thank Abdus Salam ICTP, where this work was performed, for
hospitality. A.E.F. Djema\"{i} would also like to address his
special thanks to Arab Fund and the Associateship scheme of ICTP
for their support and help. The authors are also very indebted to
A. Smailagic and E. Spallucci for useful discussions and for having shown
us their work \cite{10}.

\end{document}